%% file: main.tex
\documentclass[conference]{IEEEtran}
\IEEEoverridecommandlockouts

\usepackage{paralist}
\usepackage{xspace}
\usepackage{graphicx}
\usepackage{color}
\usepackage{fancyvrb}
\usepackage{tgcursor}

\usepackage[table,x11names]{xcolor}
\def\BibTeX{{\rm B\kern-.05em{\sc i\kern-.025em b}\kern-.08em
    T\kern-.1667em\lower.7ex\hbox{E}\kern-.125emX}}

\usepackage{url}
\usepackage{multicol}
\usepackage{booktabs}
\usepackage{ifthen}
\usepackage{subcaption}
\newcommand{\chck}[1]{#1}

\newif\ifEV
\EVtrue

\input{macros}

\begin{document}
\title{\perun: Performance Version System

% \thanks{The
% work was supported by the project 20-07487S of the Czech Science Foundation and
% the Brno Ph.D. Talent Scholarship Programme.}
}

\author{

  \IEEEauthorblockN{Tom\'{a}\v{s} Fiedor\IEEEauthorrefmark{1}, Ji\v{r}\'{\i}
  Pavela\IEEEauthorrefmark{1}, Adam Rogalewicz\IEEEauthorrefmark{1}, and
  Tom\'{a}\v{s} Vojnar\IEEEauthorrefmark{1}}

  \IEEEauthorblockA{\IEEEauthorrefmark{1}Brno University of Technology, Faculty
  of Information Technology, Brno, Czech~Republic
  % \\ifiedortom@fit.vutbr.cz, ipavela@fit.vutbr.cz, rogalew@fit.vutbr.cz, vojnar@fit.vutbr.cz
  }

}

\maketitle

\begin{abstract}
In this paper, we present \perun: an open-source tool suite for profiling-based
performance analysis.
%
% Change related to REMARK rev1.1
%
At its core, \perun maintains links between project versions and the
corresponding stored performance profiles, which are then leveraged for
automated detection of performance changes in new project versions.
The \perun tool suite further includes multiple profilers (and is designed such
that further profilers can be easily added), a performance fuzz-tester for
workload generation, 
%
% a suite of profiling optimisations,
%
methods for deriving performance models, and numerous visualization methods.
We demonstrate how \perun can help developers to analyze their program
performance on two \chck{examples}: detection and localization of a performance
degradation and generation of inputs forcing performance issues to show up.
%
% Perun repository: \url{https://github.com/tfiedor/perun/}.
%
% Alternatively my fork with the not-yet-merged code Perun repository:
% \url{https://github.com/JiriPavela/perun}
%
% Demonstration video and replication package:
% \url{https://www.fit.vutbr.cz/research/groups/verifit/tools/perun-demo/}.

\emph{Supplementary materials}---git repository~\cite{Perun}, demo
video and replication package~\cite{PerunWeb}%
\ifEV
.
\else
, and a longer version of the paper~\cite{Perun-TR}.
\fi
\end{abstract}
 

%%%%%%%%%%%%%%%%%%%%%%%%%%%%%%%%%%%%%%%%%%%%%%%%%%%%%%%%%%%%%%%%%%%%%%%%%%%%%%%
\section{Introduction}\label{sec:intro}
%%%%%%%%%%%%%%%%%%%%%%%%%%%%%%%%%%%%%%%%%%%%%%%%%%%%%%%%%%%%%%%%%%%%%%%%%%%%%%%

\ifEV
  \begin{figure*}[t]
    \centering
    \includegraphics[width=0.95\linewidth]{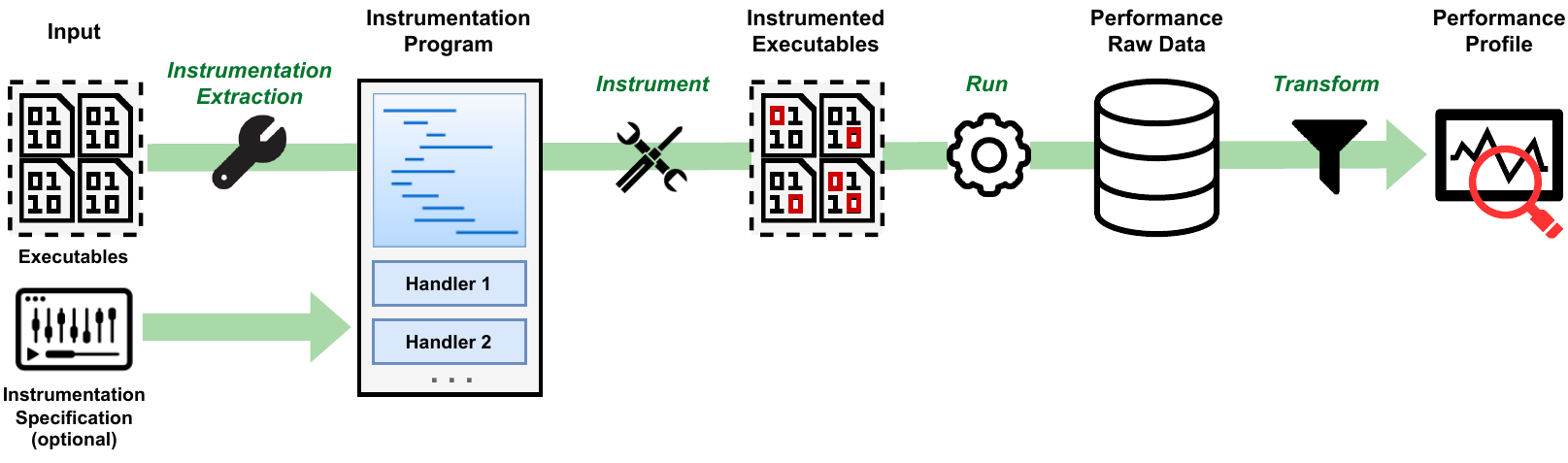}
    \caption{
      The workflow of the \tracer profiler can be divided into four
      steps. (1) The input executables are first used to identify potential
      instrumentation points (unless the user specifies them manually)
      and an instrumentation program and instrumentation handlers are generated.
      The instrumentation program instructs the instrumentation tool how to
      perform the instrumentation, and the handlers say what should be done at
      various instrumentation points. (2) The input executables are instrumented using the
      generated instrumentation program and handlers. (3) The obtained
      executable(s) are started (in a way specified by the user) and generate
      raw performance data. (4) Finally, the raw performance data are
      transformed into a performance profile.}
    %\vspace{-5.0mm}
    \label{fig:tracer-workflow}
  \end{figure*}
\else
\fi 

Performance bugs are a~common problem in software development, often encountered
during new software releases.
Despite such bugs need not crash the software, neglecting them can lead to
losing the customers' trust.
Moreover, the need of efficient software is nowadays becoming more important
than ever due to the rising scale of software and 
growing pressure on saving resources such as, e.g., battery power on mobile
devices~\cite{mobile-battery-life} or the huge power consumption of computing
centers~\cite{energy-data-centers}.
%
% Change related to REMARK rev2.5
%
Finally, software performance is also a~matter of security as performance bugs
can be exploited to effectively kill a system by some form of a
denial-of-service attack (DoS), making it run exceptionally long on some inputs.

Profiling-based analysis is arguably one of the most widely adopted techniques
used when analysing software performance.
Here, \emph{profiling} means monitoring a~running program and collecting
performance-related data, e.g., function call counts, function run times, etc.
\ifEV
  These then need to be analysed and possibly compared with previous profiling
  results to identify the most costly functions (often called \emph{hotspots} or
  \emph{bottlenecks}), to see whether some performance degradation did happen,
  and if so, what the root cause may be (and even if no degradation happened, what
  the main targets for further optimisation should be).
\else
  These then need to be analysed and possibly compared with previous
  profiling results to identify the most costly functions and to see if some
  performance degradation did happen and what its root cause may be.
\fi 

The described process is laborious and moreover repeating.
Indeed, before every new software release (at the latest), one should profile
the new version, analyze the results, compare them with past profiling, and
store the results for future profiling.
\ifEV
  Hence, it is highly desirable to automate the process as much as possible.
\else
  Hence, it is highly desirable to automate the process.
\fi  
%
% Changes related to REMARK rev2.3
%
However, while there exist various widely-used open-source \emph{profilers}
(e.g., \gprof or \callgrind), we are not aware of any actively-maintained
open-source \emph{solution} that would cover profiling-based performance
analysis in a \emph{complex} way.
Here, by complex, we mean solutions not focusing just on collecting
various performance metrics (on time, memory, energy%
\ifEV
\else
, etc.)
\fi
but also on other
needed steps (degradation detection, workload generation, etc.).
%
% Therefore, developers often invest considerable effort to build proprietary,
% ad-hoc tools to manage their project's performance.

\chck{\perun has been designed to address the above gap and provide developers
with automated, extensible, open-source support of the \emph{entire} process of
performance analysis.}
%
% we designed \emph{\perun: a performance version system}.
%
% The core of \perun maintains links between project versions and the
% corresponding results of profiling---i.e., \emph{performance profiles}.
% Change related to REMARK rev1.1
%
\chck{In line with this, \perun covers the following tasks: 1)}
\ifEV
  The core of \perun provides \emph{storage for performance profiles} (i.e.,
  results of profiling) for all project versions (chosen by the developer) and
  maintains links between the project versions and the corresponding profiles.
\else
  It provides \emph{storage for performance profiles} of all chosen project
  versions, maintaining links between the project versions and their profiles.
\fi
%
% Change related to REMARK rev2.4
%
\chck{2)~\perun} includes a highly configurable runtime \emph{profiler} called
\tracer capable of collecting various data about profiled C/C++ functions
\ifEV
 \footnote{In fact, \tracer can profile any executable (not just C/C++) with a
 symbol table.
 However, making sense of the profiling output for, e.g., executables compiled
 from automatically generated code may be difficult since the symbol table will
 likely contain a large number of functions unknown to the developer.}
\else
\fi
\chck{(note that the rest of \perun is language-agnostic)},
several other profilers, and a 
\ifEV
 wrapper-based
\else
\fi  
support for adding profilers on further metrics and/or languages.
%
%
% a suite of profiling optimisations,
%
\chck{3)~\perun also supports} \emph{fuzz-testing} for generating suitable
inputs for programs under performance analysis (i.e., the so-called
\emph{workloads}).
\chck{4)~Further, \perun} implements methods for deriving \emph{performance
models} from saved profiles as well multiple methods for automated detection of
\emph{performance degradations}, including creeping ones%
\ifEV
 ~(which require one to look deeper into the history of profiles than just to
 the immediate predecessor),
\else
  ,
\fi  
in new project versions.
\chck{5)~Finally, \perun also includes} various methods for \emph{visualising}
the obtained performance data and their models.

\ifEV
Below, we first describe the architecture of \perun and the methods implemented
in it, and then discuss various ways in which it can be used.
We illustrate the usefulness of \perun on two concrete tasks.
First, we show how it can be used to automatically detect and locate a
degradation between two software versions, namely a recent performance
degradation introduced in the 3.11.0a7 version of the \cpython
interpreter~\cite{CPython}.
Second, we present how \perun can be used to automatically generate inputs that
force performance issues to manifest, e.g., to test different hashing algorithms
or vulnerability against ReDoS (regex denial of service) attacks.
\else\fi

% \begin{figure*}[ht!]
% \centering
% \includegraphics[width=0.98\linewidth]{figures/excel-overview.pdf}
%         \vspace{-7.0mm}
% \caption{Overview of the typical workflow of the \perun.\cite{Perun}}
% \label{fig:overview}
% \end{figure*} 

%%%%%%%%%%%%%%%%%%%%%%%%%%%%%%%%%%%%%%%%%%%%%%%%%%%%%%%%%%%%%%%%%%%%%%%%%%%%%%%
\section{Architecture of \perun}\label{sec:architecture}
%%%%%%%%%%%%%%%%%%%%%%%%%%%%%%%%%%%%%%%%%%%%%%%%%%%%%%%%%%%%%%%%%%%%%%%%%%%%%%%

\begin{figure}[t!]
  \centering
  \includegraphics[width=0.95\linewidth]{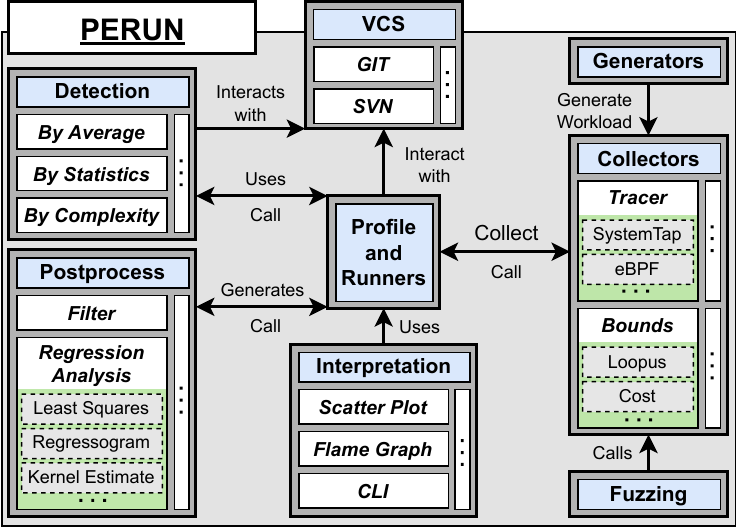}
  \vspace{-1.0mm}
  \caption{An overview of the \perun's architecture.}
  \ifEV
  \else
    \vspace{-6.0mm}
  \fi
  \label{fig:architecture}
\end{figure}

\perun consists of a~tool suite and a~wrapper over \textit{Version Control
Systems (VCS)}, such as \textit{git}, that keeps track of performance profiles
for different project versions. 
%
%Having access to both the project history and performance profiles, % \perun
%can be used to detect any potential performance changes at the moment when
%they appear: it can be set to automatically evaluate performance every time
%a~new project version is published (e.g. as a~\textit{pull-request}), and
%compares the results with the previous (stable) versions.
%
Figure~\ref{fig:architecture} shows the architecture of \perun. 
It builds on a generic JSON-based format for storing performance profiles and
other artifacts (e.g., performance models).
To implement a flexible and extensible automation of its various supported
tasks, Perun uses so-called \emph{runners} that manage \emph{jobs}: sequences
of calls of individual tools from its tool suite, i.e., data collectors,
methods deriving performance models, detection methods, or tools used for
interpretation of the results (e.g., various visualisation methods). 
Jobs may produce profiles or other results (such as performance models).
\perun's tools share a~simple API, and developers can therefore easily register
their own tools.  
Moreover most  \perun's tools are highly configurable, allowing their
fine-tuning based on domain knowledge of the profiled program. 
\perun also contains workload generators: a~fuzz-tester focused on generating
time-consuming workloads or automatic generators of gradually scaled workloads
(e.g., longer and longer files) that are fed to the profiled program.

\perun's current profilers mainly focus on profiling the runtime of functions or
programs.
The main runtime profiler is \tracer (described in more detail below).
\perun further includes a lightweight runtime profiler with minimum overhead
that is simply a wrapper over the Unix \texttt{time} utility as well as a memory
profiler, which is, however, so far experimental and not too scalable.
Other profilers (such as \callgrind) can be easily
added to \perun in a similar fashion as \texttt{time}.

Below, we describe selected tools of \perun in more detail.
%
% For more information, we refer the reader to the manual or website of
% \perun~\ref{perun}.

%---------------------------------------------------
\paragraph{Tracer} % \label{sec:tracer}
\smallskip
%---------------------------------------------------

\textit{\tracer}
% 
% (also called \textit{Trace Collector})
%
% is \perun's main profiler that 
%
% Change related to REMARK rev2.4
%
leverages multiple instrumentation frameworks to monitor programs and measure
the runtime of C/C++ functions while also keeping track of the call hierarchy
(i.e., the \textit{caller-callee} relations).
\tracer currently supports the \texttt{\systemtap}~\cite{systemtap} and the
\texttt{\ebpf}~\cite{ebpf,BPF-extension} frameworks, and can be extended by
other frameworks.
%
% (indeed, we are currently adding a support of PIN~\cite{PIN}). 
%
\ifEV
  \systemtap injects instrumentation probes
  %
  % (namely, the so-called \textit{uprobes}~\cite{UProbes} and
  % \textit{kprobes}~\cite{KProbes})
  %
  into the profiled program through an on-demand-compiled kernel module;
  while \texttt{\ebpf} applies a more dynamic approach based on a lightweight
  in-kernel virtual machine allowing better dynamic probe runtime manipulation
  (e.g., probe activation/deactivation).
\else \fi
%
% The downside of both frameworks is, however, that they require a
% custom-compiled kernel with debug information.
%
Due to its rich parameterization, \tracer is highly configurable, e.g., by
choosing the instrumentation framework, the functions to be monitored, or the
frequency with which they should be monitored.
%
% Moreover, users can also define code units other than functions to be
% profiled.

\tracer works as follows:~\begin{inparaenum}
\item It is invoked with the target executables (including libraries). The
functions to be instrumented can be specified by the user or extracted from the
executables automatically.
\item For every instrumented function, \tracer assembles and compiles
\emph{handlers} that generate the performance data.
\item It instruments the target executables, invokes the profiled program and
collects performance data until the program terminates or a~timeout is reached.
More precisely, the collected data are in the form of \emph{time stamps}
recorded at the entry and exit of each instrumented function.
%
% together with information sufficient for matching the entry/exit time stamps
% once the profiling run is over (essentially, the call context is recorded with
% the time stamps).
%
\item Finally, the raw performance data are parsed into a~profile.
\end{inparaenum}
\ifEV
  Figure~\ref{fig:tracer-workflow} illustrates the \tracer workflow.
\else
\fi

%---------------------------------------------------
\paragraph{Models and Detection Techniques} 
%---------------------------------------------------

\perun allows one to summarize profiles into \emph{performance models} mostly
having the form of functions mapping input sequence numbers to the expected
runtime.
%
% \tsf{No time now, but as we discussed, there is also other way through
% workload generators or previous tracer prototype (which required user input)}
%
% By suitably constructing the workload, the users can then get, e.g., the
% classical dependence of the runtime on the size of the input (by the workload
% exercising the program/function for larger and larger inputs), or to see how
% the performance is changing over time in a reactive system (by providing
% inputs as they would arrive in time).
%
% Change related to REMARK rev1.4
%
By suitably constructing the workload, the users can then get, e.g., the
classical \emph{complexity models} (this is, \emph{Big-O classes} such as
linear, quadratic, etc.) showing the dependence of runtime on the size of the
input (by the workload exercising the program/function for larger and larger
inputs), or to see how the performance is changing over time (by providing
inputs as they would arrive in time).
Such workloads can be generated by workload generators coming with \perun or
implemented by the users.
%
% The derived models can then be used in \perun's visualization methods and/or
% methods detecting performance degradation.
%
% Change somewhat related to REMARK rev1.4 Reformulated the sentence to be more
% clear that the models are features actively leveraged by other tools within
% Perun.
%
The derived models are then used in \perun's performance degradation detection
methods and/or visualization methods.

\perun currently supports four kinds of methods of deriving performance models:
the least-square regression line, regressogram, moving average, and kernel
regression line (for their details, see, e.g., \cite{devore, nonparam} or the
\perun's documentation).
For each function/profile we can, naturally, have multiple models.
%
% The models mostly differ in their precision, i.e., how well the model
% replicates the data from the profile.
% Change related to REMARK rev1.5
%
\perun implements multiple methods of \emph{detecting changes in the performance},
ranging from simple heuristics to more elaborate techniques, building on
well-known principles from the complexity theory~\chck{\cite{rbinfer, fb-cost, loopus}}, 
statistics~\chck{\cite{Outliers} and calculus~\cite{Integrals}}.
Each method can be applied to each pair of profiles (or models obtained from
them) for the same functions in any pair of versions of the program under
analysis.

Here are three examples of the detection methods: \begin{inparaenum}
\item The \emph{best model-order equality} builds models representing various
asymptotic complexity classes of the involved functions (linear, quadratic,
etc.), selects those that are statistically the most precise
\ifEV 
  (based on values of \emph{coefficients of determination} $R^2$)
\else
\fi
for the respective functions, and compares the orders of the models considered
the most precise for different versions of a given
\ifEV
  function (e.g., the change from a linear to a quadratic model of a
  function indicates a performance degradation).  
\else 
  function.
\fi
\item The \emph{integral comparison} approach is a heuristic based on an
assumption that the area under the performance models viewed as mathematical
functions
\ifEV
  (e.g., linear, quadratic, logarithmic functions obtained by regression
  analysis)
\else
\fi
should stay approximately the same: the technique thus computes and compares the
integrals of the most precise models.
\ifEV 
  This approach may in particular be used as a complement of the best
  model-order equality, where one sometimes gets a model of a higher order for a
  new software version but with such coefficients that make it practically the
  same as the previously derived model of a lower order (on the considered range
  of inputs). One can thus reduce the number of likely false alarms.
\else
\fi
\item The \emph{exclusive-time outliers} checks whether the
function's exclusive runtime (not including the runtime of the functions it
calls) changes so much it becomes an outlier w.r.t. other
\ifEV
  functions. In particular, this method utilizes three well-known statistical
  techniques for outlier detection using well-established thresholds for
  comparison: the modified z-score $\mathcal{Z}$ ($>3$), interquartile range $IQR$
  ($>1.5$), and standard deviation $\delta$ ($>2$).
\else
  functions based on a selected statistic (the z-score or interquartile range).
\fi
\end{inparaenum}

%---------------------------------------------------
\paragraph{Performance Fuzzing} 
%---------------------------------------------------

\perun's tool suite includes \perunfuzz: an automated fuzz-testing generator of
workloads focused on triggering performance issues. 
Fuzz-testing is an approach of automated testing that feeds a program with
randomised or mutated inputs to force unexpected/untested behaviour.
However, most of the existing approaches focus only on manifesting program
crashes, memory leaks, or failed assertions; not on manifesting performance
issues.

Well-established fuzzers, such as \afl \cite{afl-tech-details}, often build on
mutation rules that work on the level of bits in the binary encoding of program
inputs, for which, in our experience, it may be difficult to derive inputs that
could make performance issues to surface.
For \perunfuzz, we developed mutation rules inspired by various known
performance issues such as the following ones developed specifically for
text-file inputs: \fuzzrule{double the size of a line} (inspired by the issues of
the \gedit editor with long lines), \fuzzrule{repeat a random word on a line}
(to force issues, e.g., in hash tables), \fuzzrule{sort words or numbers on a
line} (to force issues in algorithms expecting randomly sorted inputs), or
\fuzzrule{prepend whitespaces} (inspired by the issues of
Stack Overflow~\cite{stackoverflow-regex} that froze this web site).

%\begin{inparaenum}
%%
%  \item \emph{Text file rules}: e.g. \fuzzrule{double the size of a line}
%  (inspired by issues of the \gedit editor with long lines), \fuzzrule{repeat a
%  random word on a line} (to force issues, e.g., in hash tables), \fuzzrule{sort
%  words or numbers on a line} (to force issues in algorithms expecting randomly
%  sorted inputs), \fuzzrule{prepend whitespaces} (inspired by issues of Stack
%  Overflow that froze the site). 
%%
%  \item \emph{Binary file rules}: e.g. \fuzzrule{remove a random zero byte},
%  \fuzzrule{add a zero byte to a random position}.
%%
%  \item \emph{Domain specific rules}: e.g. \fuzzrule{remove a~tag from an HTML
%  or XML file}.
%%
%\end{inparaenum}

Whenever a~new workload is derived, the \gcov tool that measures code coverage
(and counts also multiple visits of a~location) is used for a quick check
whether the mutation led to a coverage increased by more than some threshold.
If so, \perun is used to profile the program and to check whether some
performance degradation did indeed happen and where.

%---------------------------------------------------
\paragraph{\chck{Limitations}}
%---------------------------------------------------

\chck{\perun currently has the below limitations:
\begin{inparaenum} \item It relies mostly on dynamic analysis, hence its
analyses are limited to seen runtime behaviour.
\item Detectable performance changes are limited to those trackable by VCS;
changes not reflected in VCS, such as values of environment variables or
system/hardware settings, are not covered.
\item \perun's profiler suite is limited; it needs to be extended to support
profiling of, e.g., memory or energy consumption, and/or other languages than
C/C++.
\item Finally, \perun is limited to Linux systems, with \tracer needing a
sufficiently recent kernel.
\item \perunfuzz is limited to text-file inputs.
\end{inparaenum} }

%%%%%%%%%%%%%%%%%%%%%%%%%%%%%%%%%%%%%%%%%%%%%%%%%%%%%%%%%%%%%%%%%%%%%%%%%%%%%%%
\section{Usage of \perun}\label{sec:usage}
%%%%%%%%%%%%%%%%%%%%%%%%%%%%%%%%%%%%%%%%%%%%%%%%%%%%%%%%%%%%%%%%%%%%%%%%%%%%%%%

The most common usage of \perun is to analyze newly released program versions.
The usual scenario goes as follows:
%
%and is summarized in the Figure~\ref{fig:overview}.
%
\begin{inparaenum}
  \item The user makes a set of code changes and commits them into the
  repository of the program of interest.
  %
  % A new commit (or version) of the project is thus created.
  %
  \item The commit triggers a new build of the program, and \tracer is used over
  the resulting binary to collect new performance profiles of the involved
  functions.
  \item New performance models of the functions are derived and stored.
  \item Profiles and models of the functions from the previous version are
  retrieved from the persistent storage of \perun.
  \item For each pair of corresponding previous and new profiles (or performance
  models, depending on the detection method used), a selected performance
  degradation detection method is run.
  \item The method returns a set of located performance changes with their
  \emph{severity} (i.e., how good or bad the change is), \emph{location} (i.e.
  where the change has happened), and \emph{confidence} (i.e., how likely the
  change~is~real).
\end{inparaenum}

Another common usage scenario for \perun is in \emph{debugging}.
In its case, one can use the rich parameterization of \tracer to switch between
quick and thorough profiling while, e.g., trying to locate some suspected
bottlenecks or opportunities for optimisation.
To help this process, some of the workload generators (generating, e.g., random
files) or the performance fuzzer \perunfuzz may be used.
Further, to help understanding what happens with performance in a given program,
various well-established \emph{visualizations} of profiles and models
implemented in \perun can be used.
These include, e.g., scatter plots, bar plots, flamegraphs, or flowgraphs%
\ifEV%
---with several more in the works (for some samples, see the appendix or
\cite{PerunWeb}).
\else%
~(for some samples, see \cite{PerunWeb}).
\fi

\perun can be used through its rich command-line interface, but it can also be
easily integrated with \emph{git hooks} or other \emph{continuous integration}
solutions so that checks for performance degradation are triggered automatically
upon a commit.
Moreover, adapters based on OSLC \cite{oslc} and the \unite and \unic tools
\cite{unite-gitlab, unic-gitlab} are available for using \perun as a
%
% local or remote
%
\emph{web service} and for its integration with the \emph{Eclipse IDE}.

%%%% SHORTENING TIPS: leave out the last paragraph above.

%%%%%%%%%%%%%%%%%%%%%%%%%%%%%%%%%%%%%%%%%%%%%%%%%%%%%%%%%%%%%%%%%%%%%%%%%%%%%%%
\section{\chck{Illustrative Examples}}\label{sec:experiments}
%%%%%%%%%%%%%%%%%%%%%%%%%%%%%%%%%%%%%%%%%%%%%%%%%%%%%%%%%%%%%%%%%%%%%%%%%%%%%%%

% OLD VERSION
% We will demonstrate, how \perun and its tool suite addresses the three main
% challenges we listed in the introduction: (1) the management of projects'
% versions through the history, (2) efficient profiling and analysis through
% everyday development, and (3) inferring interesting workloads for future
% analysis.

% NEW VERSION
We now demonstrate how \perun and its tool suite can aid developers with two
common performance analysis tasks.
%
% NEW VERSION ALTERNATIVE We will demonstrate, how \perun and its tool suite can
% aid developers in % \begin{inparaenum} \item automatically detecting and
% locating performance degradation in a~new project release, and \item inferring
% interesting workloads for performance analysis.  \end{inparaenum}

%---------------------------------------------------
\paragraph{Finding Performance Degradations}
%---------------------------------------------------

First, we will focus on showing how \perun can help with identifying a
performance degradation and its root cause.
In particular, we will illustrate this on the \cpython project where a recent
release of the ver. 3.11.0a7 introduced a performance degradation compared to
the ver. 3.10.4.
The problem appeared~in the \ctypes
module\footnote{\url{https://github.com/python/cpython/issues/92356}}.
The developer who discovered~the~bug~has already proposed a fix, which has
been merged into the upstream.
Below, we replicate the issue and detect it with~\perun.

\cpython comes with a~benchmarking suite \pyperformance that is used to evaluate
the performance of various modules.
%
% on real-world use scenarios.
%
However, this particular issue has not been detected by these benchmarks.
Since the bug was reported and fixed, a new benchmark targeting the \ctypes
module has been implemented and is currently pending as a~pull request at the
time of writing of this paper.
Although the new benchmark now retrospectively detects the presence of the
\ctypes performance degradation, finding the exact source of the issue can still
pose a challenge as it requires a manual inspection of all relevant code changes
since the last version.

\perun and its version-sensitive performance analysis can nicely complement the
benchmarking approach by automatically flagging the offending function(s), thus
aiding the developers in promptly hotfixing the issue. 
\ifEV

  Our experiments with \cpython were performed on a machine with an Intel
  i7-4770 CPU at 3.4~GHz with 4 cores, 32~GiB RAM running Linux Fedora 34 with
  the x86-64 architecture.
  The system uses a custom-compiled version of the kernel 5.11.15-300.
  The custom kernel is necessary to change the default linux instrumentation
  limit of \emph{uprobes nestedness}\footnote{See the corresponding commit
  introducing the change:
  \url{https://github.com/torvalds/linux/commit/ded49c55309a37129dc30a5f0e85b8a64e5c1716}}
  (used by both \systemtap and \ebpf).
  The limit is generally acceptable for small to medium scale programs, however,
  it is too restrictive for large-scale applications, such as \cpython.
  Furthermore, each \perun profile consists of $20$ benchmark repetitions
  preceded by $80$ warm-up benchmark runs, using the default configuration as
  set in the \pyperformance benchmark\footnote{See the benchmark PR:
  \url{https://github.com/python/pyperformance/pull/197}}.

\else \fi
%
% In our case study, we did the following steps simulating the usual development
% process.
%
% Change related to REMARK rev2.9
%
% \rev{In our case study, we used the following development process.}
%
In our experiment, we did the following steps, simulating what the
developers could have done with \perun when going between the mentioned
versions:
\begin{inparaenum} \item We collected a \emph{baseline profile} of the \ctypes
benchmark, linked to the 3.10.4 release (normally, it would already be stored in
\perun).
  \item For the \cpython version 3.11.0a7, we ran \texttt{perun collect} with
  the same profiling configuration of the \ctypes benchmark as for the 3.10.4
  release and obtained a \emph{target profile} for the release 3.11.0a7.
  \item Next, we ran \texttt{perun check profiles} to perform an automated
  detection of performance changes between the \emph{baseline} and \emph{target}
  profiles. We chose the \emph{exclusive time outliers} method for detecting the
  degradation.
  \item The check successfully identified the two functions that were
  responsible for the performance degradation as shown in
  \ifEV 
    Table~\ref{tab:cs1-perfbug}.
    % Extended version
    \begin{table}[h!]
      \caption{Performance regression details of \ctypes as
      reported by the \emph{exclusive time outliers}. The two functions
    responsible for the majority of the slowdown are highlighted.
      }
      \label{tab:cs1-perfbug}
      \centering
        \begin{tabular}{l r r r} 
          \toprule
          \textbf{Location} & \textbf{Result} & \textbf{$\Delta$ [ms]} &
            \textbf{$\Delta$ [\%]}\\
          \midrule
          \rowcolor{Red1!35}
          \emph{\_ctypes\_init\_fielddesc} & NotInBaseline & 77.95 & 5.23 \\
          \rowcolor{Red1!35} \emph{\_ctypes\_get\_fielddesc} & SevereDegradation &
            52.9 & 3.55 \\
          \emph{\_ctypes\_callproc} & Degradation & 2.84 & 0.19 \\
          \multicolumn{4}{c}{\ldots} \\
          \emph{\_ctypes.cpython-311} & TotalDegradation & 136.92 & 9.19 \\
          \bottomrule
        \end{tabular}
      \end{table}
\else 
    the below table:

    % Short version
    \begin{table}[h!]
      \vspace{-2.5mm}
      % \caption{}\label{tab:cs1-perfbug} 
      % \vspace{-2.0mm}
      \centering \begin{tabular}{l| r | r r}
        \hline
          \textbf{Location} & \textbf{Result} & \textbf{$\Delta$ [ms]} & \textbf{$\Delta$
            [\%]}\\ 
        \hline 
          \rowcolor{Red1!35} \emph{\_ctypes\_init\_fielddesc} & NotInBaseline & 77.95 & 
            5.23 \\
          \rowcolor{Red1!35} \emph{\_ctypes\_get\_fielddesc} & SevereDegradation & 52.9 & 
            3.55 \\
          \emph{\_ctypes\_callproc} & Degradation & 2.84 & 0.19 \\
          \multicolumn{4}{c}{\ldots} \\ 
          \emph{\_ctypes.cpython-311} & TotalDegradation & 136.92 & 9.19 \\ 
        \hline
      \end{tabular}
      \vspace{-2.5mm}
    \end{table}
  \fi
  \noindent The $\Delta [ms]$ ($\Delta [\%]$) is the absolute (relative)
  exclusive time change w.r.t. the total duration of the program or library
  which contains the function.
  %
  % Change related to REMARK rev2.9
  %
  Note that the total degradation we measured is comparable to the
  degradation of around 8\,\% reported in the original issue.
  \item We created a hotfix in a new VCS branch \texttt{hotfix-issue-92357} and
  repeated the profiling to obtain a new \emph{target profile} for the branch
  \texttt{hotfix-issue-92357}.
  \item Finally, we checked that the fix was successful: we ran \texttt{perun
  check profiles} using the original \emph{baseline} and the new \emph{target}
  profiles. This time, the check report
  \ifEV
    given in Table~\ref{tab:cs1-fixed}
  \else
  \fi
  shows that both of the previously degraded functions have comparable
  performance with the release 3.10.4 (see the table below). Note that the
  \ctypes module still reports a minor degradation, corresponding to the
  \pyperformance benchmarking results.
  %
  % \item Alternatively, the developers can run just the \ctypes benchmark on
  % the \texttt{hotfix-issue-92357} branch to confirm the performance
  % degradation has been resolved.  
  %
\end{inparaenum}

\ifEV
  % Extended version
  \begin{table}[h!]
    \caption{
      The result of the \emph{exclusive time outliers} detection method after the
      hotfix. The previously degraded functions are now showing a negligible
      change of their duration only.
    }\label{tab:cs1-fixed}
    \centering
    \begin{tabular}{l r r r}    
      \toprule
        \textbf{Location}  & \textbf{Result} & \textbf{$\Delta$ [ms]} & 
          \textbf{$\Delta$ [\%]} \\ 
      \midrule
        \emph{\_ctypes\_callproc} & SevereDegradation & 9.7 & 0.7 \\
        \multicolumn{4}{c}{\ldots} \\
        \rowcolor{Green1!35} \emph{\_ctypes\_get\_fielddesc} & MaybeDegradation & 0.89 & 
          0.06 \\
        \rowcolor{Green1!35} \emph{\_ctypes\_init\_fielddesc} & NotInBaseline & 0.02 &
          0.00 \\
        \emph{\_ctypes.cpython-311} & TotalDegradation & 23.45 & 1.70 \\
      \bottomrule
    \end{tabular}
  \end{table}
\else
  % Short version
  \begin{table}[h!]
    % \vspace{-1.5mm}
    % \caption{}\label{tab:cs1-fixed}
    \vspace{-2.0mm}
    \centering
    \begin{tabular}{l| r | r r}    
      \hline
        \textbf{Location}  & \textbf{Result} & \textbf{$\Delta$ [ms]} & 
          \textbf{$\Delta$ [\%]} \\ 
      \hline
        \emph{\_ctypes\_callproc} & SevereDegradation & 9.7 & 0.7 \\
        \multicolumn{4}{c}{\ldots} \\
        \rowcolor{Green1!35} \emph{\_ctypes\_get\_fielddesc} & MaybeDegradation & 0.89 & 
          0.06 \\
        \rowcolor{Green1!35} \emph{\_ctypes\_init\_fielddesc} & NotInBaseline & 0.02 &
          0.00 \\
        \emph{\_ctypes.cpython-311} & TotalDegradation & 23.45 & 1.70 \\
      \hline
    \end{tabular}
    \vspace{-3mm}
  \end{table}
\fi

In the usual profiling scenario, should the developer wish to continue with
optimizations, the next most severe degradation would be targeted next.
The \emph{exclusive time outliers} method aids the developer with this effort
and reports a new severe degradation, relative to the other found degradations:
\emph{\_ctypes\_callproc}.
To define a stopping point for this workflow, a \emph{cut-off} threshold for
$\Delta$ can be set, so that no degradations or improvements below this
threshold are reported.

For detection of \emph{creeping performance degradations}, one can utilize the
same approach as above with a slight modification only.
Namely, the \emph{baseline} profile should be selected such that it corresponds
to an older project version or release. 

% \tsf{Missing discussion (short) of the results} \jp{Resolved, please check if
% it is enough}

% We note that the profiling can be optimized by measuring only the functions
% present in the binaries \texttt{python3.10},
% \texttt{\_ctypes.cpython-310-x86\_64-linux-gnu.so}, and their counterparts
% for version 3.11.0a7.

% TsF: Probably too much inf o
%The \emph{exclusive time outliers} method detects when a function
%\emph{exclusive} time (i.e., time spent only in the actual function, without
%the time spent in the callee functions) has changed so much that it becomes an
%outlier.
%%
%More precisely, the method employs three different outlier detection algorithms
%based on the well-established \emph{modified z-score}, \emph{IQR multiple} and
%\emph{standard deviation multiple} statistical techniques.

%---------------------------------------------------
\paragraph{Inferring Interesting Workloads}
%---------------------------------------------------

In our second \chck{example}, we will demonstrate how to automatically generate
workloads suitable for performance analysis. In this \chck{example}, we will use
minimalistic single-purpose projects. We believe that such programs are enough
at least if our goal is to find workloads for testing specific code patterns. 
%
% Naturally, such generated workloads are usable for larger projects as well.

In particular, we will generate workloads to 
%
% (1) data structures whose performance depend on their balanced height; 
%
(1) identify potentially expensive regex matching and (2) to compare the
efficiency of various hash functions used to implement a hash table.
\ifEV
  All experiments with inferring the interesting workloads were run on a Lenovo
  G580 machine with an Intel Core i3-3110M CPU at 2.40~GHz using 4 cores, 4~GiB
  RAM, and Ubuntu 18.04.2 LTS.
\else
\fi
For each experiment, we list the size of each workload denoted as $w_i$ (with
the starting, manually prepared workload denoted as $seed$), the runtime of the
program under test on the given workload, the slowdown w.r.t. the runtime on the
seed, and the number of mutation rules that led to the workload.

% First, we generated workloads to test the Unbalanced Binary Tree.
% Theoretically, its insertion should be in $\mathcal{O}(n.log(n))$, however,
% when the inputs are sorted it degrades to $\mathcal{O}(n^2)$ complexity. As a
% corpus we used files each with 10,000 randomly generated integers.
% Table~\ref{tab:cs2-ubt} shows the results. Using initial seeds generates tree
% with height 21 and takes about 0.1s; the best workload generated by our fuzzer
% using a single rule leads to a tree height between 3,000-6,000 and performance
% degradation of several orders of magnitude using.

% \begin{table}[h!]
% \caption{}\label{tab:cs2-ubt}
% \vspace{-2.0mm}
% \centering
% \begin{tabular}{l| r r r r}
%     \multicolumn{5}{c}{Unbalanced Binary Tree}\\
% \hline
%     \textbf{input}  & \textbf{size [B]} & \textbf{time [s]} & \textbf{slowdown} & \textbf{used rules}\\ 
% \hline 
% \hline
% \emph{seed} & 48,913 & 0.109 & - & - \\
% \hline
% \emph{$w_1$} & 49,912 & 11.014 & 101.04 & 1\\
% \hline
% \end{tabular}
%         \vspace{-6.0mm}
% \end{table}

First, we focus on regex matching that can be very expensive for some matchers
and some regexes, potentially admitting a ReDoS (Regex Denial of Service)
attack~\cite{owasp-regex, stackoverflow-regex, email-regex}.
These were reported in various applications, e.g., the~following regex used for
Java classname validation~\cite{owasp-regex}:
\verb#^(([a-z])+.)+[A-Z]([a-z])+$#.
The issue here is that the developer did not escape the `\texttt{.}'
character, which makes some matchers to perform an excessive number of
backtracking steps on some non-matching words.
%
% Carefully matching an input then leads to a denial of a~service.
%
% Note that there exists other (more sophisiticated) approaches for generating
% inputs for such patterns~\cite{redos1, redos2}, however, we wish to
% demonstrate that our fuzzer is generic enough to cover such a~case.

We tested the \texttt{std::regex\_search} function with several offending regexes
that can cause ReDoS attacks. 
\ifEV
  In Table~\ref{tab:cs2-regex},
\else
  Below,
\fi
we present results for the above Java Classname
validation regex.
%
% from OWASP Validation Regex Repository~\cite{owasp-regex}.
%
For other regexes and results, see \cite{PerunWeb}.

\ifEV
  % Extended version
  \begin{table}[h!]
    \caption{
      An illustration of the \perun's workload generator discovering problematic
      inputs, potentially leading to a ReDoS on the Java Classname validation
      regex, in only a few applications of mutation rules.
    }\label{tab:cs2-regex}
    \centering
    \begin{tabular}{l r r r r}
    \toprule
      \multicolumn{5}{c}{\textbf{Java Class validation regex}~\cite{owasp-regex}}\\
      \textbf{Input}  & \textbf{Size [B]} & \textbf{Time [s]} & \textbf{Slowdown} & 
        \textbf{Used rules}\\ 
    \midrule
      \emph{seed}       & 19 & 0.005 & - & - \\
    \midrule
      \emph{$w_2$} & 19 & 0.016 & 3.2 & 9 \\
      \emph{$w_3$} & 36 & 1.587 & 317.4 & 6 \\
      \rowcolor{Red1!35} \emph{$w_5$} & 78 & \textbf{timeout} (13h) & $\infty$ & 5 \\
    \bottomrule
    \end{tabular}
  \end{table}
\else
  % Short version
  \begin{table}[h!]
    \vspace{-2.5mm}
    % \caption{}\label{tab:cs2-regex}
    % \vspace{-2.0mm}
    \centering
    \begin{tabular}{l| r r r r}
      \multicolumn{5}{c}{Java Class validation regex~\cite{owasp-regex}}\\
    \hline
      \textbf{input}  & \textbf{size [B]} & \textbf{time [s]} & \textbf{slowdown} & 
        \textbf{used rules}\\ 
    \hline 
    \hline
      \emph{seed}       & 19 & 0.005 & - & - \\
    \hline
      \emph{$w_2$} & 19 & 0.016 & 3.2 & 9 \\
      \emph{$w_3$} & 36 & 1.587 & 317.4 & 6 \\
      \rowcolor{Red1!35} \emph{$w_5$} & 78 & \textbf{timeout} (13h) & $\infty$ & 5 \\
    \hline
    \end{tabular}
    \vspace{-2.5mm}
  \end{table}
\fi

\ifEV
  The results in Table~\ref{tab:cs2-regex} show that, 
\else
  \noindent Clearly,
\fi
at least for the considered regexes, we can find
\ifEV
  workloads possibly triggering ReDoS attacks
\else
  problematic workloads
\fi
with only a few applications of our mutation rules.
Indeed, some generated inputs were only 4x bigger than the seed, yet leading to
a timeout of thirteen hours.

Finally, we tested a simple word frequency counter that uses a hash table.
We compared its performance using two different hash functions: (1) the hash
function used in the Java 1.1 string library, which examined only 8--9 evenly
spaced characters, 
%
% which, however, has a~high collision rate for long strings;
%
and, (2) the DJB hash function~\cite{DJB}, one of the most efficient hash
functions.
The results
\ifEV
  in Table~\ref{tab:cs2-hash}
\else
\fi
show that with increasing sizes of the workload, the DJB hash
function does indeed perform in a much more stable way compared to the inefficient
implementation of Java 1.1.

\ifEV
  % Extended version
  \begin{table}[h!]
    \caption{
      Performance comparison of two hashing algorithms using the \perun's
      workload generator. The Java 1.1 hashing algorithm is shown to scale
      less w.r.t larger inputs.
    }\label{tab:cs2-hash}
    \centering
    \begin{tabular}{l r | r r | r r}
      \toprule    
        \multicolumn{2}{c|}{\textbf{Work}} & \multicolumn{2}{c|}{\textbf{Java 1.1.}} & \multicolumn{2}{c}{\textbf{DJB}}\\
        \textbf{Input}  & \textbf{Size [kB]} & \textbf{Time [s]} & \textbf{Slowdown} & \textbf{Time [s]} & \textbf{Slowdown} \\ 
      \midrule
        \emph{seed} & 210 & 0.026 & - & 0.013 & - \\
      \midrule
        \emph{$w_6$} & 458 & 0.115 & 4.4 & 0.027 & 2.1 \\
        \emph{$w_7$} & 979 & \cellcolor{Red1!35} 0.187 & \cellcolor{Red1!35} \textbf{7.19} & 0.043 & 3.3\\
      \bottomrule
    \end{tabular}
  \end{table}
\else
  % Short version
  \begin{table}[h!]
    \vspace{-2.5mm}
    % \caption{}\label{tab:cs2-hash}
    % \vspace{-2.0mm}
    \centering
    \begin{tabular}{l| r | r r | r r}    
        \multicolumn{2}{c}{Work} & \multicolumn{2}{c}{Java 1.1.} & \multicolumn{2}{c}{DJB}\\
        \hline
            \textbf{input}  & \textbf{size [kB]} & \textbf{time [s]} & \textbf{slowdown} & \textbf{time [s]} & \textbf{slowdown} \\ 
    \hline
            \emph{seed} & 210 & 0.026 & - & 0.013 & - \\
    \hline
            \emph{$w_6$} & 458 & 0.115 & 4.4 & 0.027 & 2.1 \\
            \emph{$w_7$} & 979 & \cellcolor{Red1!35} 0.187 & \cellcolor{Red1!35} \textbf{7.19} & 0.043 & 3.3\\
    \hline
    \end{tabular}
            \vspace{-4.0mm}
  \end{table}
\fi

%%%%%%%%%%%%%%%%%%%%%%%%%%%%%%%%%%%%%%%%%%%%%%%%%%%%%%%%%%%%%%%%%%%%%%%%%%%%%%%
\section{Related Work}\label{sec:related}
%%%%%%%%%%%%%%%%%%%%%%%%%%%%%%%%%%%%%%%%%%%%%%%%%%%%%%%%%%%%%%%%%%%%%%%%%%%%%%%

% Change related to REMARK rev2.11
%
Perhaps the closest solution to \perun is \perfrepo~\cite{PerfRepo} whose idea
is similar to what \perun aims to achieve, but it covers profiling-based
analysis in a less complex way (e.g., it does not include a profiler, workload
generation, etc.) and it lacks a tight VCS integration (this is, it lacks
built-in linkage between profiles and project VCS versions).
Moreover, its development seems to be discontinued, based on the GitHub
repository activity.
\kieker~\cite{kieker} provides a tool suite for performance data collection and
results interpretation but lacks any form of project version context.
\valgrind~\cite{Nethercote-Valgrind}, a~widely-adopted profiling tool suite,
lacks support for results comparison, and VCS/CI integration.

A range of individual profilers, either sampling or event-based, for heap,
memory or CPU profiling exist. We can list, e.g., \oprofile~\cite{OProfile},
\perf~\cite{perf-tool}, \gprof~\cite{gprof}, or \gperftools~\cite{gperftools}.
Furthermore, we can mention various commercial solutions such as, e.g., the
\emph{Intel V-Tune Profiler}~\cite{VTune}, \emph{AMD $\mu$Prof}~\cite{MuProf},
\emph{Arm MAP}~\cite{ArmMAP}, or the \emph{Oracle Developer Studio Performance
Analyzer}~\cite{OraclePerformanceAnalyzer}.
Additionally, there are other performance analysis tools based on machine
learning~\cite{perf-anal-ml}, code patterns~\cite{perf-anal-similar-access},
learning from performance bug trackers~\cite{perf-anal-repo-extract}, automated
and compositional load test
generation~\cite{automatic-load-gen,compositional-load-gen}, checking
correlation between performance testing results~\cite{perf-anal-mining},
clustering and comparing performance counters~\cite{perf-anal-clusters}, or
using statistical process control techniques~\cite{perf-anal-control}.
However, these tools do either not provide as complex coverage of
profiling-based performance analysis as \perun does (being often restricted to
profiling only) and/or they are not open source, often quite expensive and/or
coming with some additional hardware or software requirements for unlocking
their full potential.

Finally, as for performance fuzzing, the only tools of this kind we are aware of
are \perffuzz~\cite{perffuzz} and \badger~\cite{perf-fuzz-badger}.
Unlike \perunfuzz, which is, however, just one part of \perun, they use other
fuzzing rules, working on the binary level as in \afl (\badger combines fuzzing
with symbolic execution too).
As an alternative to performance fuzzing, causal profiling~\cite{coz} aims at
finding particular functions to optimise (by slowing down other functions to
virtually simulate the speed-up). 
\perfblower~\cite{perfblower} forces memory-related errors by amplifying their
effects on performance.
A support of such techniques could be added to \perun in the future.

\ifEV

%%%%%%%%%%%%%%%%%%%%%%%%%%%%%%%%%%%%%%%%%%%%%%%%%%%%%%%%%%%%%%%%%%%%%%%%%%%%%%%
\section{Conclusion}\label{sec:conclusion}
%%%%%%%%%%%%%%%%%%%%%%%%%%%%%%%%%%%%%%%%%%%%%%%%%%%%%%%%%%%%%%%%%%%%%%%%%%%%%%%

\perun is a new, open-source framework for managing project's performance
throughout its development history and for debugging performance-related issues.
It links performance profiles to their respective project versions to keep the
context of performance changes, and, moreover, comes with a comprehensive suite
of performance tools that collect, analyze, model and visualize performance
data.
As for the future, we are working on further performance data collectors,
visualisation methods, as well as optimisations of the profiling process.
We plan to use various statically or dynamically collected pieces of information
to leave out functions with minimal performance impact from the profiling (or to
profile them sparsely only).
\else
\fi

% Acknowledgment
%%%%%%%%%%%%%%%%%%%%%%%%%%%%%%%%%%%%%%%%%%%%%%%%%%%%%%%%%%%%%%%%%%%%%%%%%%%%%%%
\vspace*{4mm} {\fontsize{7}{8}\selectfont \noindent \emph{Acknowledgement.} The
work was supported by the project 20-07487S of the Czech Science Foundation and
the Brno Ph.D. Talent Scholarship Programme.\par} %\eject

%%%% SHORTENING TIPS: leave out the last sentence above.
\clearpage

\bibliographystyle{IEEEtran}
\bibliography{literature}

\ifEV
  \begin{figure*}[ht] 
    \begin{subfigure}[b]{0.5\linewidth}
      \centering
      \includegraphics[width=0.95\linewidth]{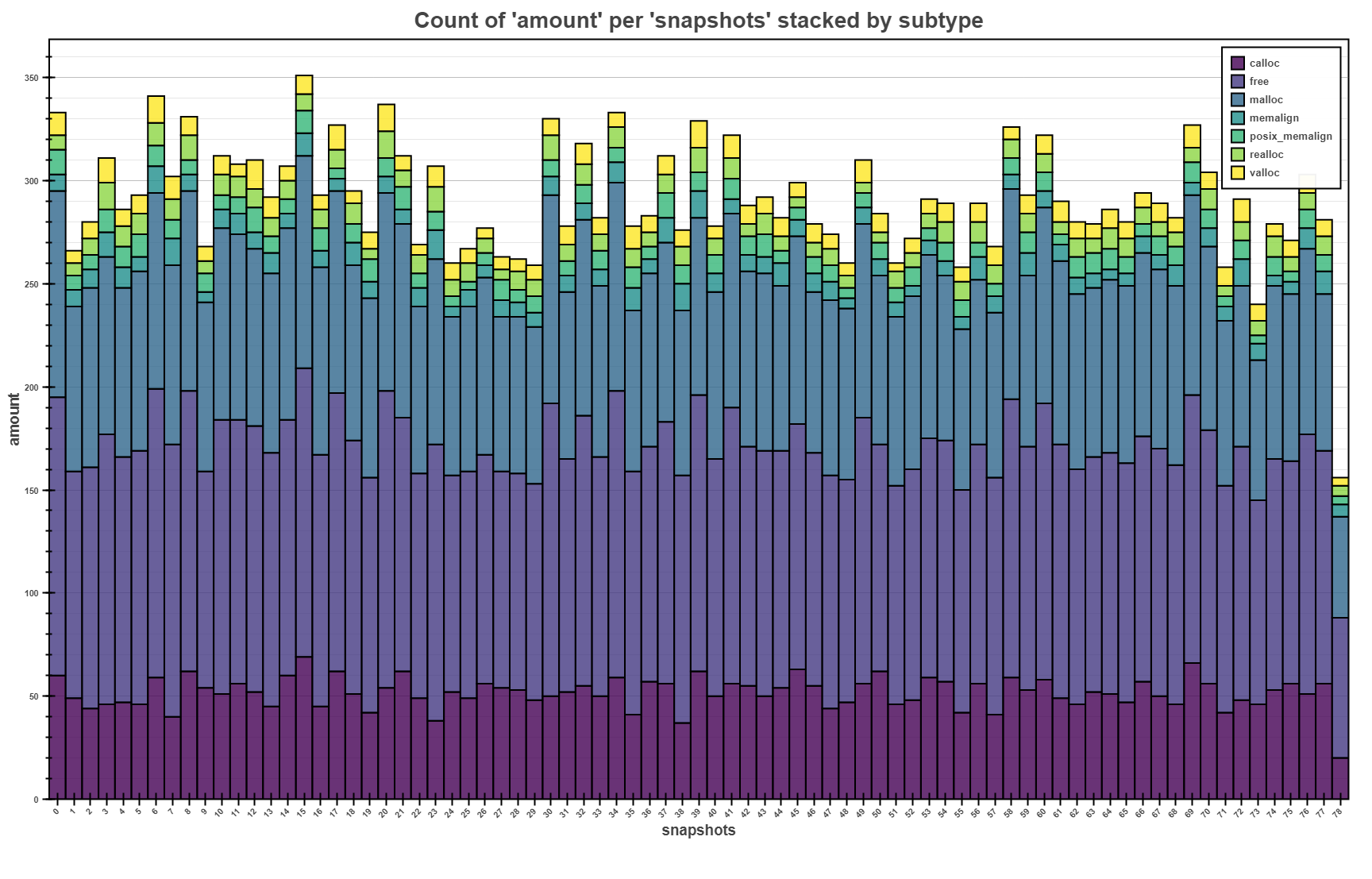} 
      \caption{An example of a stacked bar graph of memory manipulation.} 
      \label{fig:vis-bargraph}
      \vspace{2ex}
    \end{subfigure}%% 
    \begin{subfigure}[b]{0.5\linewidth}
      \centering
      \includegraphics[width=0.95\linewidth]{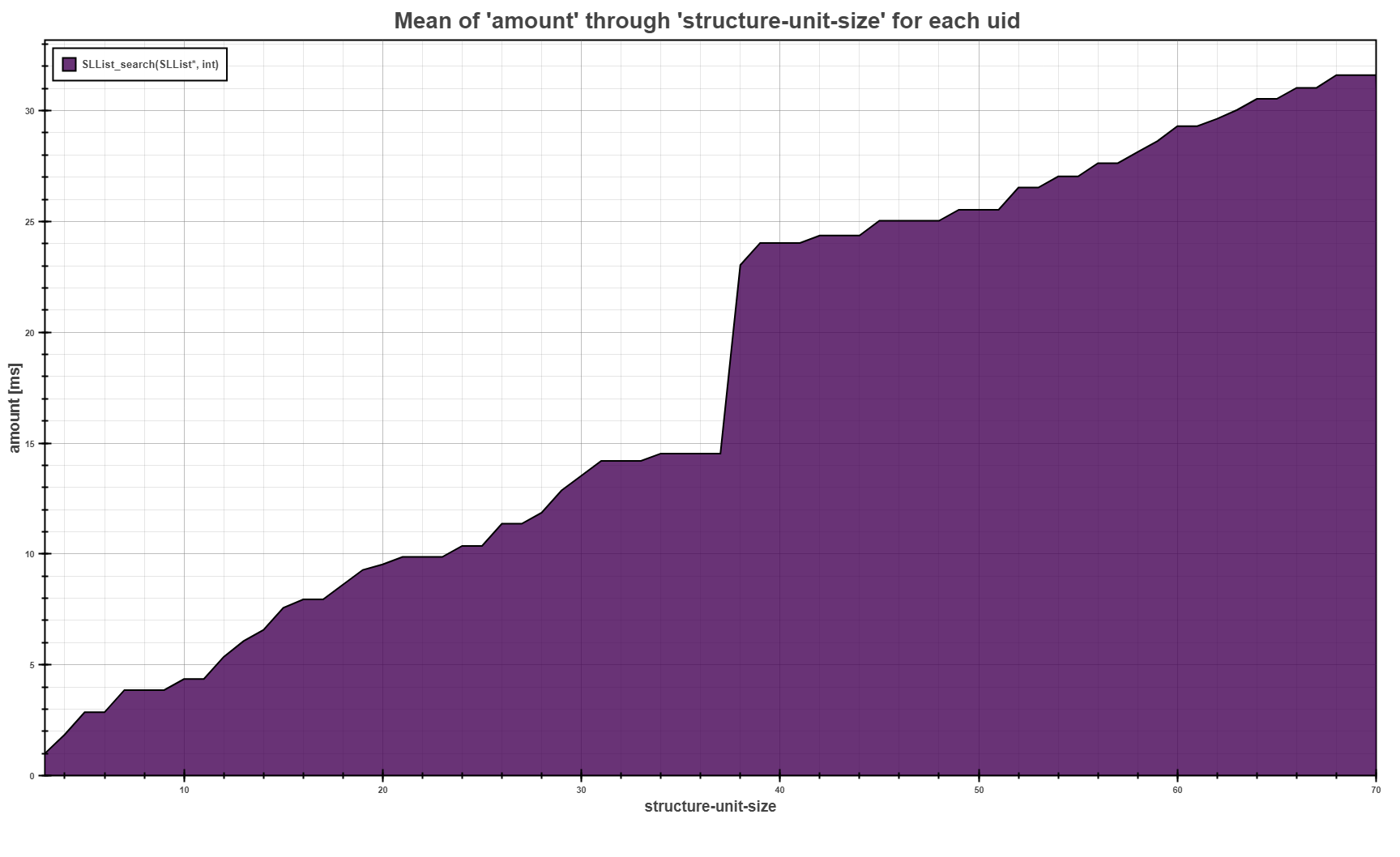} 
      \caption{An example of a time consumption flow graph.} 
      \label{fig:vis-flowplot} 
      \vspace{2ex}
    \end{subfigure} 
    \begin{subfigure}[b]{0.5\linewidth}
      \centering
      \includegraphics[width=0.95\linewidth]{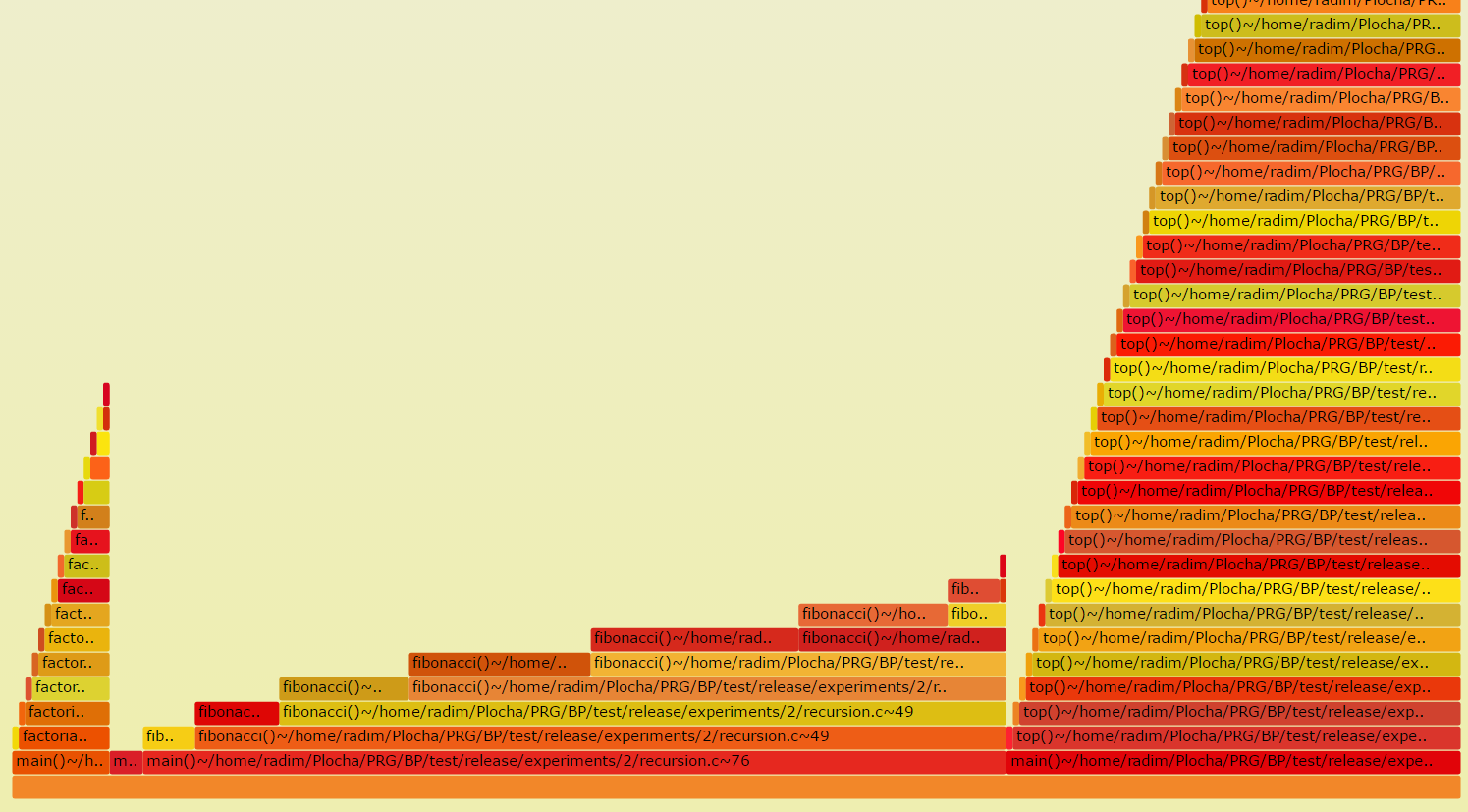} 
      \caption{An example of a memory flame graph.} 
      \label{fig:vis-flamegraph} 
    \end{subfigure}%%
    \begin{subfigure}[b]{0.5\linewidth}
      \centering
      \includegraphics[width=0.95\linewidth]{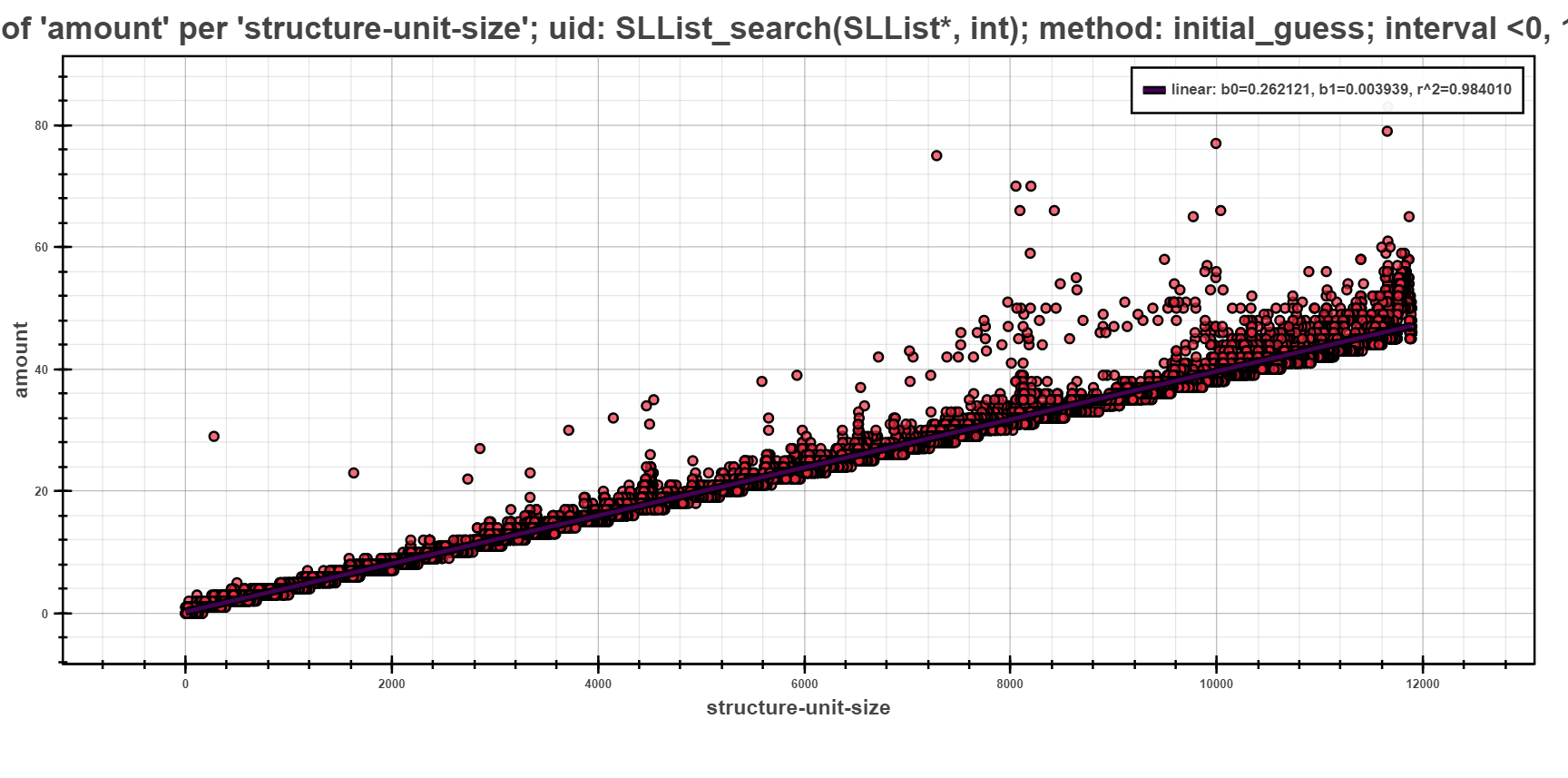} 
      \caption{An example of a scatter plot with the best regression model.} 
      \label{fig:vis-scatterplot} 
    \end{subfigure} 
    \caption{Examples of visualization methods implemented by \perun.}
    \label{fig:visgrid} 
    %\vspace*{-4mm}
  \end{figure*}

\newpage

\appendices

%%%%%%%%%%%%%%%%%%%%%%%%%%%%%%%%%%%%%%%%%%%%%%%%%%%%%%%%%%%%%%%%%%%%%%%%%%%%%%%
\section{Visualization Examples}\label{appendix}
%%%%%%%%%%%%%%%%%%%%%%%%%%%%%%%%%%%%%%%%%%%%%%%%%%%%%%%%%%%%%%%%%%%%%%%%%%%%%%%

\perun supports a number of visualization methods, namely \emph{bar plots},
\emph{flow plots}, \emph{flame graphs} and \emph{scatter plots}.
In general, some visualizations may be applicable, or suitable, only for
certain profile types.
Below, we briefly discuss some of the supported visualisation means.
To see the full gallery of Perun visualization capabilities, please visit the
\perun tool page~\cite{PerunWeb}.

%%%%%%%%%%%%%%%%%%%%%%%%%%%%%%%%%%%%%%%%%%%%%%%%%%%%%%%%%%%%%%%%%%%%%%%%%%%%%%%
\subsection{Bar Plots}
%%%%%%%%%%%%%%%%%%%%%%%%%%%%%%%%%%%%%%%%%%%%%%%%%%%%%%%%%%%%%%%%%%%%%%%%%%%%%%%

Bar plots depict collected data about \emph{consumed resources} (e.g., runtime,
memory consumption) as bars, which can be displayed in either a stacked or
grouped format.
Figure~\ref{fig:vis-bargraph} shows an example of a stacked bar plot of memory
consumption by different functions at regular time points during the run of a
program (\emph{snapshots}).

% \begin{figure}[t]
%   \centering
%   \includegraphics[width=0.95\linewidth]{figures/visualizations/07-memory-bars-stacked.png}
%   \vspace{-3.0mm}
%   \caption{An example of a stacked bar graph.}
%   \vspace{-3.0mm}
%   \label{fig:vis-bargraph}
% \end{figure} 

%%%%%%%%%%%%%%%%%%%%%%%%%%%%%%%%%%%%%%%%%%%%%%%%%%%%%%%%%%%%%%%%%%%%%%%%%%%%%%%
\subsection{Flow Plots}
%%%%%%%%%%%%%%%%%%%%%%%%%%%%%%%%%%%%%%%%%%%%%%%%%%%%%%%%%%%%%%%%%%%%%%%%%%%%%%%

Flow plots show the dependency of the recorded resource consumption on some
independent variable chosen by the user (e.g., the size of the input, the
sequential number of the input, etc.).
Similarly to bar plots, flow plots can also accumulate or stack the resource
consumption values.
For example, Figure~\ref{fig:vis-flowplot} shows the trend of the average
duration of an \texttt{SLList\_search} function, providing some implementation
of searching in \emph{singly-linked lists} (SLLs), depending on the size of the
SLL it is executed on.

% \begin{figure}[t]
%   \centering
%   \includegraphics[width=0.95\linewidth]{figures/visualizations/05-complexity-flow.png}
%   \vspace{-3.0mm}
%   \caption{An example of a flow graph.}
%   \vspace{-3.0mm}
%   \label{fig:vis-flowgraph}
% \end{figure} 

\subsection{Flame Graphs}

Flame graphs show the relative recorded consumption of resources w.r.t. the
execution trace leading to the program construction that caused the resource
consumption.
Flame graphs aid in faster localization of resource consumption hotspots.
Figure~\ref{fig:vis-flamegraph} shows the memory consumption of a~simple program
that calls selected recursive functions, such as Fibonacci or factorial.
Each rectangle represents an allocation of the memory, and its width is scaled
according to how much memory has been allocated at a~given function and a~given
line (e.g., when we call the \texttt{malloc} function).

% \begin{figure}[t]
%   \centering
%   \includegraphics[width=0.95\linewidth]{figures/visualizations/10-memory-flamegraph.png}
%   \vspace{-3.0mm}
%   \caption{An example of a flame graph.}
%   \vspace{-3.0mm}
%   \label{fig:vis-flamegraph}
% \end{figure} 

\subsection{Scatter Plots}

Scatter plots visualize recorded resource consumption as points on a two dimensional
grid where the X-axis (Y-axis) typically represents values
of an independent variable (the collected resource value), respectively.
Figure~\ref{fig:vis-scatterplot} shows the dependency of the
\texttt{SLList\_search} function on the size of the \emph{single-linked list}
structure it is executed on.
Furthermore, the plot also shows the best regression model (based on the $R^2$
\emph{coefficient of determination} value) that captures the dependency of the
runtime on the structure size.

% \begin{figure}[t]
%   \centering
%   \includegraphics[width=0.95\linewidth]{figures/visualizations/02-view-scatter.png}
%   \vspace{-3.0mm}
%   \caption{An example of a scatter plot with the best derived model.}
%   \vspace{-3.0mm}
%   \label{fig:vis-scatterplot}
% \end{figure} 

\else
\fi

\end{document}

%% file: macros.tex
\newcommand{\cpython}{\textsc{CPython}\xspace}
\newcommand{\perun}{\textsc{Perun}\xspace}
\newcommand{\tracer}{\textsc{Tracer}\xspace}
\newcommand{\valgrind}{\textsc{Valgrind}\xspace}
\newcommand{\callgrind}{\textsc{Callgrind}\xspace}

\newcommand{\systemtap}{\textsc{SystemTap}\xspace}
\newcommand{\ebpf}{\textsc{eBPF}\xspace}

\newcommand{\kieker}{\textsc{Kieker}\xspace}
\newcommand{\badger}{\textsc{Badger}\xspace}
\newcommand{\perfblower}{\textsc{PerfBlower}\xspace}

\newcommand{\perffuzz}{\textsc{Perffuz}\xspace}
\newcommand{\perunfuzz}{\textsc{Perun-fuzz}\xspace}
\newcommand{\gedit}{\texttt{gedit}\xspace}
\newcommand{\ctypes}{\texttt{ctypes}\xspace}
\newcommand{\gprof}{GNU \texttt{gprof}\xspace}
\newcommand{\perfrepo}{\textsc{PerfRepo}\xspace}
\newcommand{\unite}{\textsc{Unite}\xspace}
\newcommand{\unic}{\textsc{Unic}\xspace}
\newcommand{\pyperformance}{\texttt{pyperformance}\xspace}

\newcommand{\oprofile}{\textsc{OProfile}\xspace}
\newcommand{\perf}{\textsc{Perf}\xspace}
\newcommand{\gperftools}{\texttt{gperftools}\xspace}
\newcommand{\afl}{\texttt{afl}\xspace}
\newcommand{\gcov}{\texttt{gcov}\xspace}

\newcommand{\fuzzrule}[1]{\emph{``#1"}}